\documentclass{article}

\usepackage{arxiv}

\usepackage[utf8]{inputenc} 
\usepackage[T1]{fontenc}    
\usepackage{hyperref}       
\usepackage{url}            
\usepackage{booktabs} 
\usepackage{multirow} 
\usepackage{siunitx} 
\usepackage{amsfonts}       
\usepackage{nicefrac}       
\usepackage{microtype}      
\usepackage{lipsum}
\usepackage{graphicx}
\usepackage{amsmath}
\usepackage{authblk}

\title{Shape-aware synthesis of pathological lung CT scans using CycleGAN for enhanced semi-supervised lung segmentation}
\author[1,4]{Rezkellah Noureddine Khiati\textsuperscript{*}}
\author[2]{Pierre-Yves Brillet}
\author[3]{Aurélien Justet}
\author[4]{Radu Ispas} 
\author[1]{Catalin Fetita \textsuperscript{*}} 

\affil[1]{SAMOVAR, Telecom Sud-Paris, Institut Polytechnique de Paris, Evry, France}

\affil[2]{Avicenne Hospital, AP-HP, Bobigny, France}
\affil[3]{Pneumologie et Médecine Nucléaire CHU Caen, France}
\affil[4]{Keyrus France, Levallois-Perret, France}

\begin{document}
\maketitle
\footnotetext[1]{Contributed equally.}
\begin{abstract}
This paper addresses the problem of pathological lung segmentation, a significant challenge in medical image analysis, particularly pronounced in cases of peripheral opacities (severe fibrosis and consolidation) because of the textural similarity between lung tissue and surrounding areas.
To overcome these challenges, this paper emphasizes the use of CycleGAN for unpaired image-to-image translation, in order to provide an augmentation method able to generate fake pathological images matching an existing ground truth. Although previous studies have employed CycleGAN, they often neglect the challenge of shape deformation, which is crucial for accurate medical image segmentation. Our work introduces an innovative strategy that incorporates additional loss functions. Specifically, it proposes an L1 loss based on the lung surrounding which shape is constrained to remain unchanged at the transition from the healthy to pathological domains. The lung surrounding is derived based on ground truth lung masks available in the healthy domain. Furthermore, preprocessing steps, such as cropping based on ribs/vertebra locations, are applied to refine the input for the CycleGAN, ensuring that the network focus on the lung region. This is essential to avoid extraneous biases, such as the zoom effect bias, which can divert attention from the main task.
The method is applied to enhance in semi-supervised manner the lung segmentation process by employing a U-Net model trained with on-the-fly data augmentation incorporating synthetic pathological tissues generated by the CycleGAN model. The combined use of CycleGAN for data augmentation and U-Net for segmentation leads to a very effective and reliable method for image segmentation requiring a reduced amount of native data for training.
Preliminary results from this research demonstrate significant qualitative and quantitative improvements, setting a new benchmark in the field of pathological lung segmentation. Our code is available at \url{https://github.com/noureddinekhiati/Semi-supervised-lung-segmentation}
\end{abstract}



\section{Introduction}
Lung segmentation is an essential task for achieving precise phenotyping and longitudinal follow-up of interstitial lung diseases. Many algorithms use markers from CT scans of the lungs, such as measurements of vessels and airways,  to generate indices that are instrumental in the early detection or diagnosis of various pathologies \cite{peng2022quantitative}, including COVID-19 and fibrosis. Several approaches have been proposed in the literature, from simple (region growing - able to deal with near-healthy subjects) to more complex (mathematical morphology \cite{inproceedings}, fuzzy-logic \cite{6850014}, etc. able to tackle different pathologies). Recently, deep learning approaches have gained interest for the segmentation tasks because the higher execution speed and model sharing opportunities. Typically, these approaches utilize region-growing algorithms to generate preliminary masks \cite{mesanovic2011automatic}, which are subsequently refined by experts to provide the ground truth for model training.  However, the region growing algorithm exhibits limitations, particularly in the presence of severe pathological alterations in the lungs due to the textural resemblance between lung tissue and adjacent anatomical structures. Certain studies, such as \cite{CHEN2021105864}, have adopted an enhanced random walker algorithm to address this challenge. Nonetheless, this technique does not achieve absolute precision and requires expert refinement, a process that is both expensive and time-consuming. 

On the other hand, some studies use a wide range of labeled data, including both (near-)healthy and diseased lungs and their masks \cite{Hofmanninger_2020}. The access to such annotated data is however not granted to the large research community, which limits the technical development.

Data augmentation is a prevalent strategy in medical imaging \cite{chlap2021review}, particularly when annotated data are scarce. In our scenario, simulating pathological tissue may prove beneficial in training segmentation models to recognize the lung in the presence of pathology. A potential solution involves the generation of synthesized pathological tissue within (near-)healthy lung scans (for which lung annotation is easier and available), to train in a semi-supervised manner a neural network, commonly a U-Net \cite{ronneberger2015unet}, for segmentation. This data augmentation approach aims to foster the network's proficiency in delineating pathological lung structures.

CycleGANs \cite{zhu2020unpaired} are notably effective in this domain, functioning as unsupervised image-to-image translators. Although CycleGANs have been applied in medical imaging, a caveat exists: for segmentation purposes, post-domain translation maintaining anatomical shape integrity is paramount. Deformation in shape may induce a mismatch with the labeled ground truth and compromise segmentation accuracy.

\cite{CONNELL2022200} have already proposed CycleGANs for pathological lung segmentation. However, these studies did not incorporate shape awareness and operated standard CycleGAN frameworks. They transformed pathological lung images into their "healthy" counterparts and subsequently applied a segmentation model designed for healthy lungs. This approach, while very appealing, presents no guarantee for lung shape preservation when translating from pathological to "healthy" domain and our implementation of this method revealed this limitation. 

This paper addresses the previous limitations and proposes a data augmentation strategy based on a robust CycleGAN generation model that emphasizes lung shape preservation. Notably, our model achieves significant results even when trained on a limited dataset, by incorporating a loss term penalising deviations from the lung shape and by focusing on the lung region using an image cropping based on the rib cage, which is proven essential for the CycleGAN framework. Similarly, for the segmentation phase, we have exploited a standard U-Net model  trained with augmented pathological data provided by the proposed CycleGAN. This combination between the generation and segmentation models ensures enhanced performance even with a small training native dataset, showcasing the potential of our integrated approach in medical imaging tasks where data availability is often a challenge.

\section{Pathology-generative model}
\subsection{Data Description} 
\label{sec:data_desc}
For the training of the CycleGAN, we manually selected a subset of 150 axial images from a COVID follow-up study (\textit{COVALUNG} dataset, Caen University Hospital, France), specifically for representing the healthy domain. This dataset is well-annotated, covering 57 patients followed-up post-COVID and having recovered from disease (near-healthy subjects), each contributing a varying number of slices throughout the lung.

In contrast, for the pathological domain, we sourced our images from a retrospective cohort of lung interstitial diseases dataset (\textit{SILICOVILUNG}, Avicenne Hospital, Bobigny, France). These selected slices exhibit severe pathology characterized by peripheral dense tissue (fibrosis, consolidation). It is important to note that this dataset is not annotated.

\subsection{Data preprocessing} 
\label{sec:data}
\begin{figure}[htbp]

   
  \includegraphics[width=\linewidth]{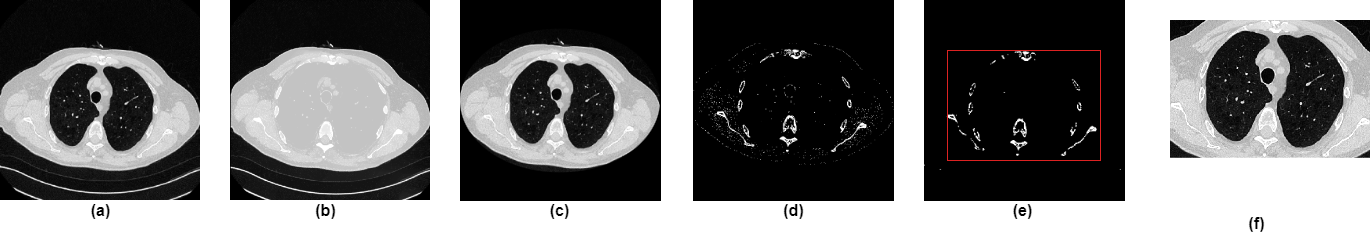}
  \caption{Detailed pipeline for rib cage-based cropping operation}
  \label{fig:pipeline_cropping}
\end{figure}

To mitigate the bias caused by the zoom-in and zoom-out effects observed in several images from both datasets, particularly during CycleGAN training where the model might learn the zoom effects instead of focusing on the primary task, we implemented a cropping operation. This operation is based on the rib cage, a common feature present in both domains – near-healthy and pathological lungs. The rib cage serves as a stable reference point, ensuring the retention of essential information about lungs.
Our preprocessing approach involves several steps, as follows: 
\begin{itemize}
    \item \textbf{Selection of the body region:} The lung area is filled-in with high intensity values using grayscale morphological reconstruction by erosion with respect to image border (Figure \ref{fig:pipeline_cropping}b). After image binarisation, the largest connected component is selected (Figure \ref{fig:pipeline_cropping}c) in order to remove high intensity structures  outside the thorax cage (eg. scanner bed). 
    \item \textbf{Rib cage segmentation:} A high level thresholding applied in the body region provides the dense structures including the rib cage. A morphological opening removes the noise outside the rib cage (Figure \ref{fig:pipeline_cropping}d). 
    \item \textbf{Image cropping:} The bounding box of the rib cage (Figure \ref{fig:pipeline_cropping}e) is computed automatically and subsequently used to crop the original image, thus emphasizing the lung region (Figure \ref{fig:pipeline_cropping}f). 
\end{itemize}
The cropped images were resized to 256x256 pixels for the CycleGAN training.

\subsection{Model architecture}
The CycleGAN architecture consists of two Generators ($G$ and $F$) and two Discriminators ($D_{x}$ and $D_{y}$) (Figure 2). Generator $G$ maps images from domain $X$ (near-healthy lung CT scans) to domain $Y$ (pathological lung scans), while Generator F maps images from domain $Y$ to domain $X$. Discriminator $D_{x}$ aims to distinguish between images from domain $X$ and generated images by F, while $D_{y}$ distinguishes between images from domain $Y$ and generated images by G.
\\
To ensure that the generated pathological lung scans retain the anatomical structure of the near-healthy lungs, we integrate an L1 loss based on the lung masks from the healthy domain. The L1 loss is computed on the region surrounding the lung mask, to emphasize structural consistency in these areas. This is particularly crucial for subsequent utilization in a U-Net model for lung segmentation, where preserving lung shape is mandatory.
   
The lung surrounding is extracted using the lung mask $M$ from the healthy domain images. Let $I_{healthy}$ be the input post-COVID lung CT scan, and $I_{generated}$ be the generated pathological scan by the generator $G$. The surrounding lung region $S$ is obtained by subtracting the lung mask from the full image area, focusing the L1 loss on the areas outside the lung mask:
\begin{equation}\label{eq:L1_loss}
    S_{healthy} = (1-M) \odot I_{healthy}
\end{equation}
The $L1$ loss is then applied between the corresponding regions of $S_{healthy}$ and $S_{generated}$: 
\begin{equation}\label{eq:example}
\mathcal{L}_{L1}(S_{healthy},S_{generated} ) = \frac{1}{N} \sum_{i=1}^{N} | S_{healthy}(i) - S_{generated}(i) | 
\end{equation}
where $S_{generated} = (1 - M) \odot I_{generated}$  represents the lung surrounding region in the generated pathological image, $N$ is the number of pixels in the region of interest, 
$| \cdot | $ denotes the absolute value, and the operation is applied element-wise to the difference of the two inputs.

The overall loss function of the CycleGAN with integrated L1 loss is formulated as follows: 
\begin{equation}\label{eq:Total_loss}
\begin{aligned}
    \mathcal{L}_{total} = & \mathcal{L}_{GAN}(G, D_Y, X, Y) + \mathcal{L}_{GAN}(F, D_X, Y, X) + \delta \mathcal{L}_{cyc}(G, F)  +\lambda \mathcal{L}_{L1}(G, M)  \\
    & +\gamma \mathcal{L}_{identity}(G, Y)  +\gamma \mathcal{L}_{identity}(F, X)
\end{aligned}
\end{equation}

where $\mathcal{L}_{GAN}(G, D_Y, X, Y)$  and  $\mathcal{L}_{GAN}(F, D_X, Y, X)$ represent the adversarial losses for the mapping functions G and F, respectively, $\mathcal{L}_{cyc}(G, F)$  is the cycle consistency loss that enforces  $F(G(x)) \approx x$  and  $G(F(y)) \approx y$, $\mathcal{L}_{identity}$ is the identity loss that forces $F(x) \approx x$ and $G(y) \approx y$   and $\delta$  and  $\lambda$  are hyperparameters controlling the relative importance of the cycle consistency loss and the L1 loss, respectively.

In the development of our model, the choice of loss functions for the generators and discriminators was guided by the foundational principles laid out in the original CycleGAN paper. Specifically, we opted for the Mean Squared Error (MSE) loss to quantify the adversarial losses \( \mathcal{L}_{GAN}(G, D_Y, X, Y) \) and \( \mathcal{L}_{GAN}(F, D_X, Y, X) \). This choice aligns with the standard approach in CycleGAN frameworks, where the $MSE$ loss has been demonstrated to effectively measure the discrepancy between the distribution of generated images and real images.

\begin{figure}[htbp]
  
  \includegraphics[width=\linewidth]{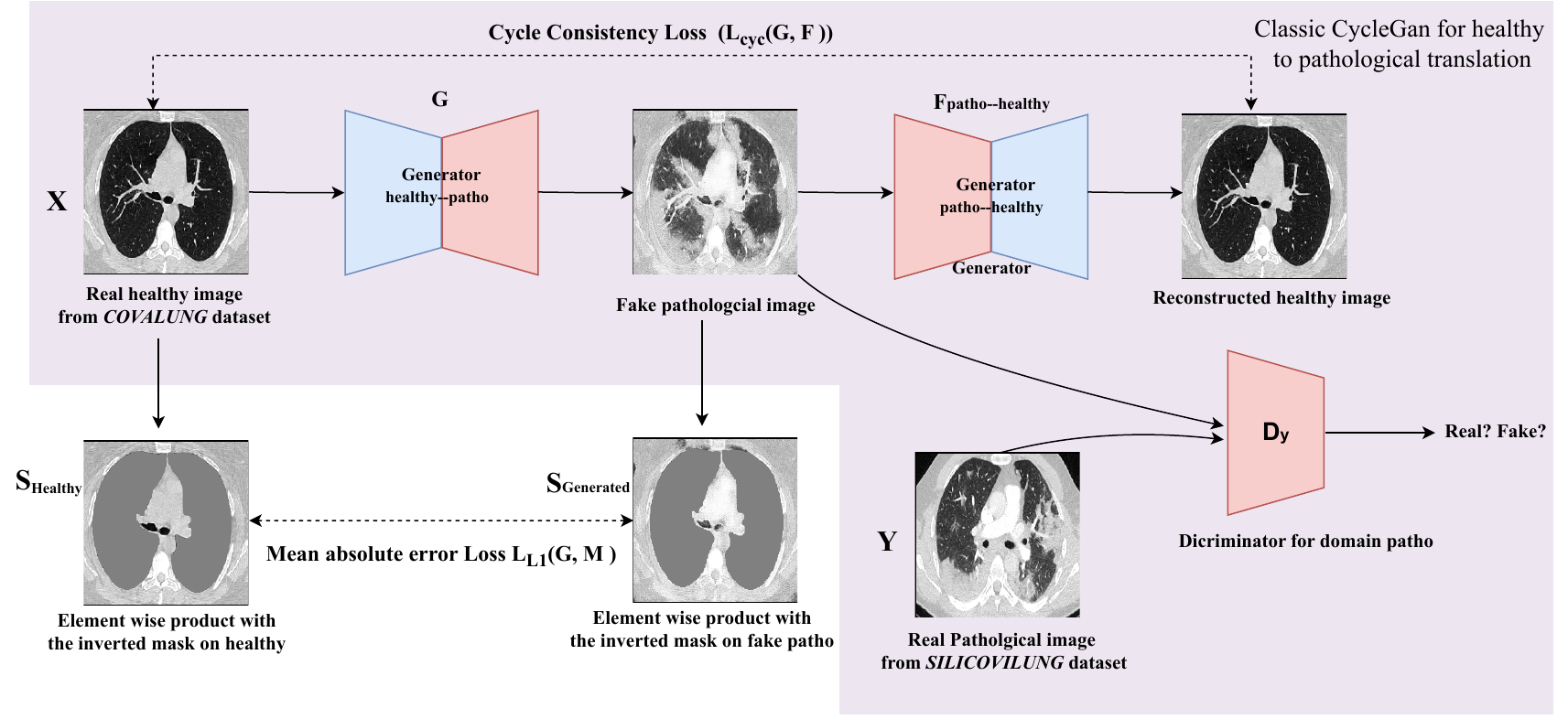}
  \caption{Architecture of the modified CycleGAN with L1 loss integration. Only the translation healthy to pathological is shown (the reverse translation implies reverting $X$ and $Y$, $G$ and $F$ and replacing $D_{y}$ by $D_{x}$).}  
  \label{fig:pipeline_cycle}
\end{figure}

For the cycle consistency loss \( \mathcal{L}_{cyc}(G, F) \), we adhered to the recommendation in the original CycleGAN paper and employed the L1 loss. The complete pipeline of our modified CycleGAN, incorporating the L1 loss for improved synthesis of pathological lung CT scans, is shown in Figure \ref{fig:pipeline_cycle}.

However, our empirical observations indicated that incorporating the identity loss did not contribute positively to the early learning phases of our network. Specifically, we noticed that the model, when guided by the identity loss, tended to leave the images virtually unchanged, impeding the transformation from the healthy to the pathological domain. Consequently, the identity loss was dismissed from our model. 

\begin{figure}[htbp]

  \includegraphics[width=1\linewidth]{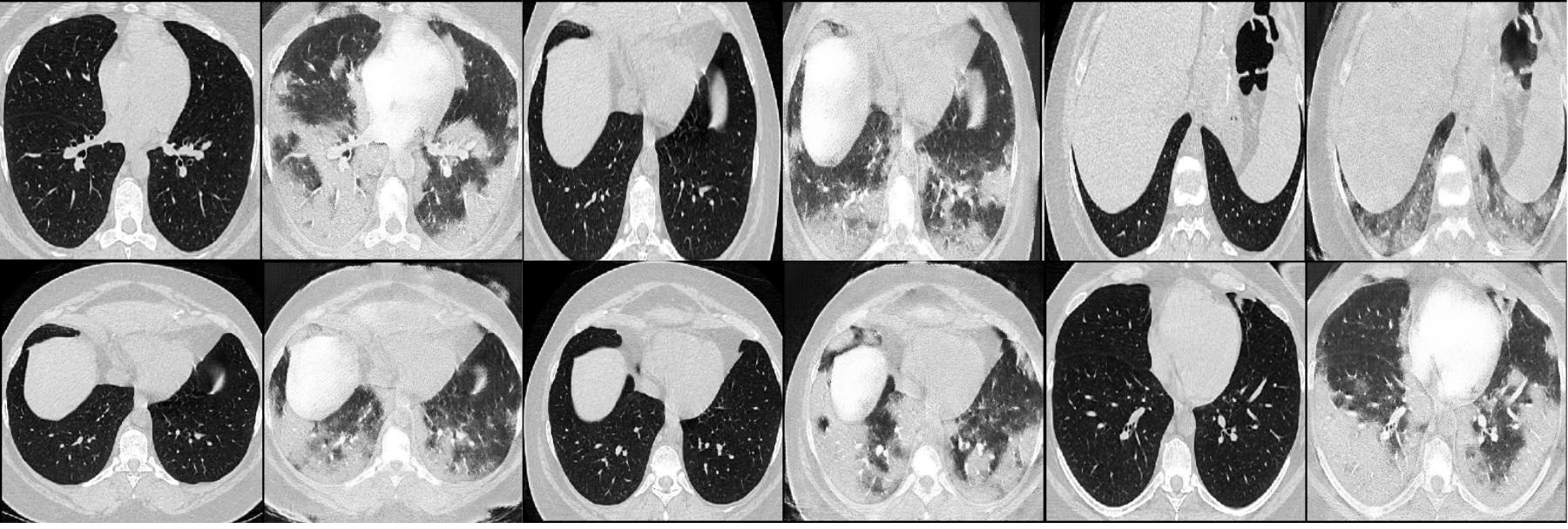}
  \caption{Qualitative results from the test set display paired images, showcasing the transformation from healthy to pathological domains.}
  \label{fig:post-COVID2patho}
\end{figure}

We trained our CycleGAN model on cropped images from the dataset described in (§\ref{sec:data_desc}). The training process spanned 100 epochs to ensure adequate model convergence and learning. Post-training, the model performance was evaluated on a separate test set comprising 500 images from each domain. The evaluation was primarily qualitative, focusing on the visual and structural accuracy of the generated images in reflecting the respective domain characteristics (Figure \ref{fig:post-COVID2patho}).

In terms of hyperparameters, after a thorough empirical study and optimization, we set the cycle consistency weight \( \delta \) to 10. Furthermore, the weight \( \lambda \) for the L1 loss was fixed at 1. This parameter was fine-tuned to ensure that the model adequately preserves the anatomical structures surrounding the lung region, as delineated by the L1 loss component. A detailed description of the model optimization is presented in Appendix \ref{appendix:c}. 


The model demonstrates commendable performance in translating from the healthy to pathological domain while preserving the lung shape, as illustrated in Figure \ref{fig:post-COVID2patho}. This augmentation will be used later in our segmentation model to enhance its accuracy. However, in the reverse translation from pathological to healthy, the absence of a shape preservation mechanism becomes obvious since no constraints could be set in this respect. As shown in Appendix \ref{appendix:a} (Figure \ref{fig:pathological_to_post-COVID}), the model exhibits inaccuracies in cleansing the lung from pathologies, which prevents using this feature in the segmentation phase.

\section{Segmentation model} 
For the segmentation task, we utilized the near-healthy \textit{COVALUNG} dataset as training database, which comprises CT scans from a total of 57 patients, each patient having a varying number of annotated slices with a maximum of 200 slices per patient. To maintain a focus on lung structures, we retained only 5\% of non-lung slices for each patient. The dataset was split based on individual patients into training, validation, and test sets with proportions of 0.7, 0.1, and 0.2, respectively.

In terms of the network architecture, we used   the original U-Net framework, ensuring the same number of features at each level of the network. Enhancements to the model included the incorporation of batch normalization after each convolutional block to stabilize and accelerate training. Additionally, a dropout rate of 0.2 was introduced in the decoder blocks to mitigate overfitting.

All images were preprocessed by cropping on the rib cage (see §\ref{sec:data}), maintaining consistency with the preprocessing performed for the CycleGAN. Our trained CycleGAN model was employed as a means of data augmentation, introducing synthesized pathological lung CT images into the training process. On-the-fly data augmentation was applied with a probability of 0.5. Furthermore, we introduced Gaussian blur augmentation with a kernel size of \(5 \times 5\) immediately after the application of the CycleGAN-generated images. This augmentation was aimed at increasing the difficulty of the segmentation task, encouraging the model to learn more nuanced features, especially in terms of the lung contour gradients, which are not visible  in case of the original pathological images.

The model was trained using the Dice-Focal loss. The focal component of the loss, with a gamma parameter set to 2, was specifically designed to address the class imbalance issue by focusing more on hard-to-segment regions.
To further enhance the robustness of the model and prevent overfitting, which was observed to occur in the early epochs, we implemented a cyclic annealing learning rate scheduler in conjunction with the SGD optimizer. The model underwent training for a total of 100 epochs, leveraging this dynamic adjustment of the learning rate to navigate the complex optimization landscape effectively.

\section{Results and discussion}
A preliminary qualitative evaluation on pathological images from the test \textit{SILICOVILUNG} dataset showed pertinent segmentation results (Appendix \ref{appendix:a} Figure \ref{fig:segmentation_silicovilung}). For a quantitative assessment, we employed different test sets with available annotations, as follows:
\begin{itemize}
\item \textbf{COVID-2019 challenge dataset:} We analyzed data from 33 patients exhibiting severe COVID-19 symptoms with lung consolidation. Note that the available ground truth in this dataset consists solely of Covid area masks. To obtain complete lung masks, we employed an additional model known for its reliability in normal lung condition segmentation.
\item \textbf{Severe pathology subset:} A focused evaluation on 333 axial slices from the COVID-2019 Challenge Dataset, showcasing severe consolidation patterns.
\item \textbf{ILD168 dataset:} We extended our analysis to 17 patients from an in-house dataset including patients with infiltrative lung diseases (ILD), highlighting less severe pathologies like Fibrosis, GGO, and Emphysema.
\item \textbf{Coronacases Radiopedia dataset:} The first 10 patients from this public dataset \cite{MP-COVID-19-SegBenchmark} were also included to diversify the pathologies assessed.
\end{itemize}
We compare our model with several baseline approaches. The models evaluated were the  following: 
\begin{itemize}
\item \textbf{U-Net post-COVID:} A baseline UNet model trained exclusively on the post-COVID (\textit{COVALUNG}) dataset (40 patients) without synthetic pathology data augmentation. 
\item \textbf{U-Net extended:} An extended UNet model incorporating training data from both the post-COVID dataset and additional 117 patients (70\% of the ILD168 dataset), focusing on a wider array of lung pathologies.
\item \textbf{V-Net:} In order evaluate a 3D model and observe the effect of the extra added dimension, we evaluated a V-Net architecture trained on the post-COVID training set, with a patch size of 64. It is important to note that for testing, we only utilized data from Radiopaedia and the COVID-2019 challenge, since only these sources provided the full 3D datasets required for our analysis.
\item \textbf{CycleGAN-UNet:} Our proposed U-Net model, utilizing CycleGAN for data augmentation, trained on the post-COVID dataset.
\item \textbf{R231 Model:} A benchmark model known for its effectiveness in pathological lung segmentation \cite{Hofmanninger_2020}, sourced from the COVID website and integrated into the 3D Slicer software. R231 is trained on a proprietary dataset including 231 patients with various lung pathologies.
\end{itemize}

 The average scores are summarized in the table below, with per-patient details (Appendix \ref{appendix:b}) and some illustrations in Appendix \ref{appendix:a} (Figure \ref{fig:results_segmentation_covid}).

\begin{table}[htbp]
\centering
\label{tab:dice_scores_comparison}
\resizebox{\textwidth}{!}{%
\begin{tabular}{
    l
    S[table-format=3.0]
    S[table-format=3.0]
    S[table-format=2.2]
    S[table-format=2.2]
    S[table-format=2.2]
    S[table-format=2.2]
    S[table-format=2.2]
}
\toprule
& \multicolumn{2}{c}{Dataset details} & \multicolumn{5}{c}{Models} \\
\cmidrule(lr){2-3} \cmidrule(lr){4-8}
{Dataset} & {Slices} & {Patients} & {U-Net post-COVID} & {U-Net Extended} & {V-Net} & {R-231} & {CycleGAN+U-Net} \\
\midrule
ILD168 & 251 & 17 & 84.78 & 96.93 & / & \textbf{98.19} & 97.98 \\
Radiopedia & 2581 & 10 & 83.67 & 85.55 & 84.35 & \textbf{98.29} & 97.77  \\
Subset-333 & 333 & 33 & 80.23 & 83.16 & / & 91.88 & \textbf{93.07}  \\
COVID challenge & 2219 & 33 & 74.78 & 88.60 & 86.20 & 93.69 & \textbf{93.91}  \\
\bottomrule
\end{tabular}%
}
\caption{Dice scores comparison of models on different datasets}

\end{table}

The results highlight the competitive performance of our CycleGAN + U-Net model in comparison with the other models. Notably, our model achieves a comparative Dice scores with the state of the art R231 model, indicating its effectiveness in pathological lung segmentation, even in the absence of real pathological tissue in the initial training dataset.
\section{Concluding remarks}
It is worth noting that our CycleGAN + U-Net model was trained on a relatively small dataset, comprising only 40 patients for the U-Net training in addition to the 150 slices used for the CycleGAN. In contrast, the R231 model was trained on a significantly larger dataset, including 231 patients with annotated pathological lungs. Despite the smaller training set, our model demonstrates competitive performance, underscoring the effectiveness of the data augmentation provided by the CycleGAN-generated images. This suggests that the proposed training strategy improves model generalization capacity, even when trained on a limited amount of native data.

This study was funded by the \textit{Fondation du Souffle}, France (under the COVALung project) and by the French National Research Agency (under the ANR MLQ-CT project).
\newpage
\bibliographystyle{unsrt} 
\bibliography{references}

\newpage
\appendix

\section{Qualitative analysis}
\label{appendix:a}
\subsection{Qualitative analysis of CycleGAN translations from pathological images to post-COVID}
We present a qualitative evaluation of the results obtained from applying the CycleGAN model to translate pathological lung CT images of \textit{SILICOVILUNG } test set into their healthy version. In this situation, lung masks are not available and shape-preservation constraints cannot be incorporated into the model. The results demonstrate the limitation of this type of domain translation and the importance to add the shape constraint.

\begin{figure}[htbp]
\centering
\includegraphics[width=1\linewidth]{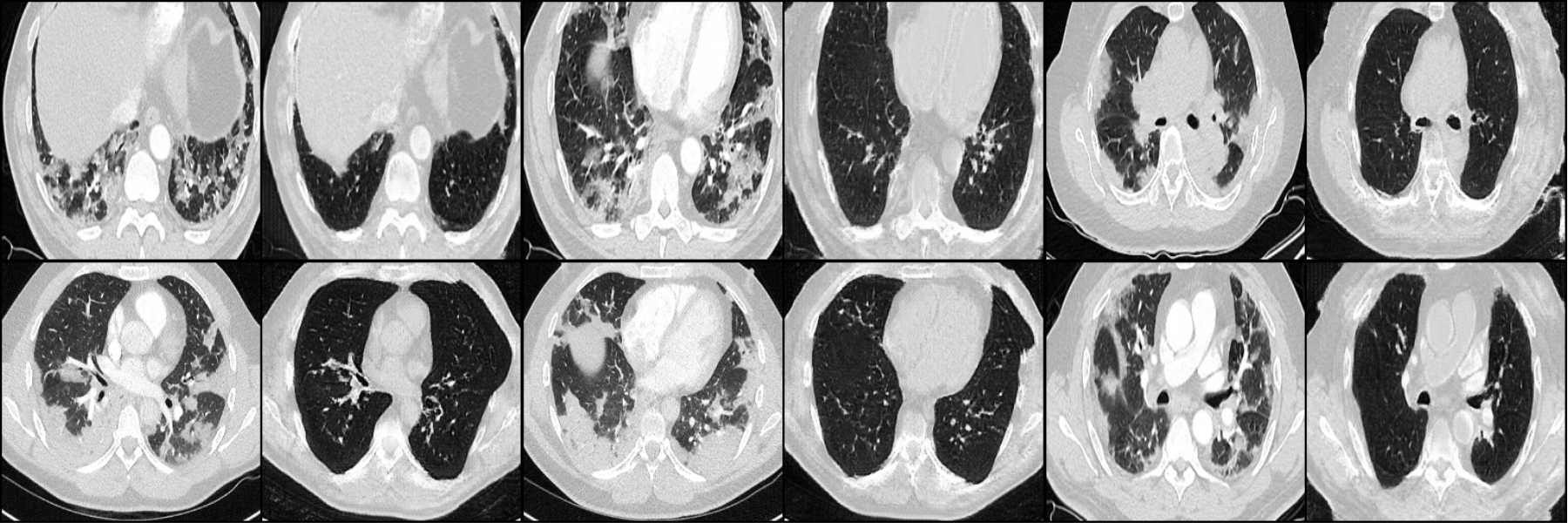}

\caption{Qualitative results from the test set display paired images, showcasing the transformation from pathological to post-COVID domains.}
\label{fig:pathological_to_post-COVID}
\end{figure}

\subsection{Qualitative segmentation results on SILICOVILUNG test set}
\begin{figure}[htbp]
\vspace{-5pt} 
\centering
\includegraphics[width=1\linewidth]{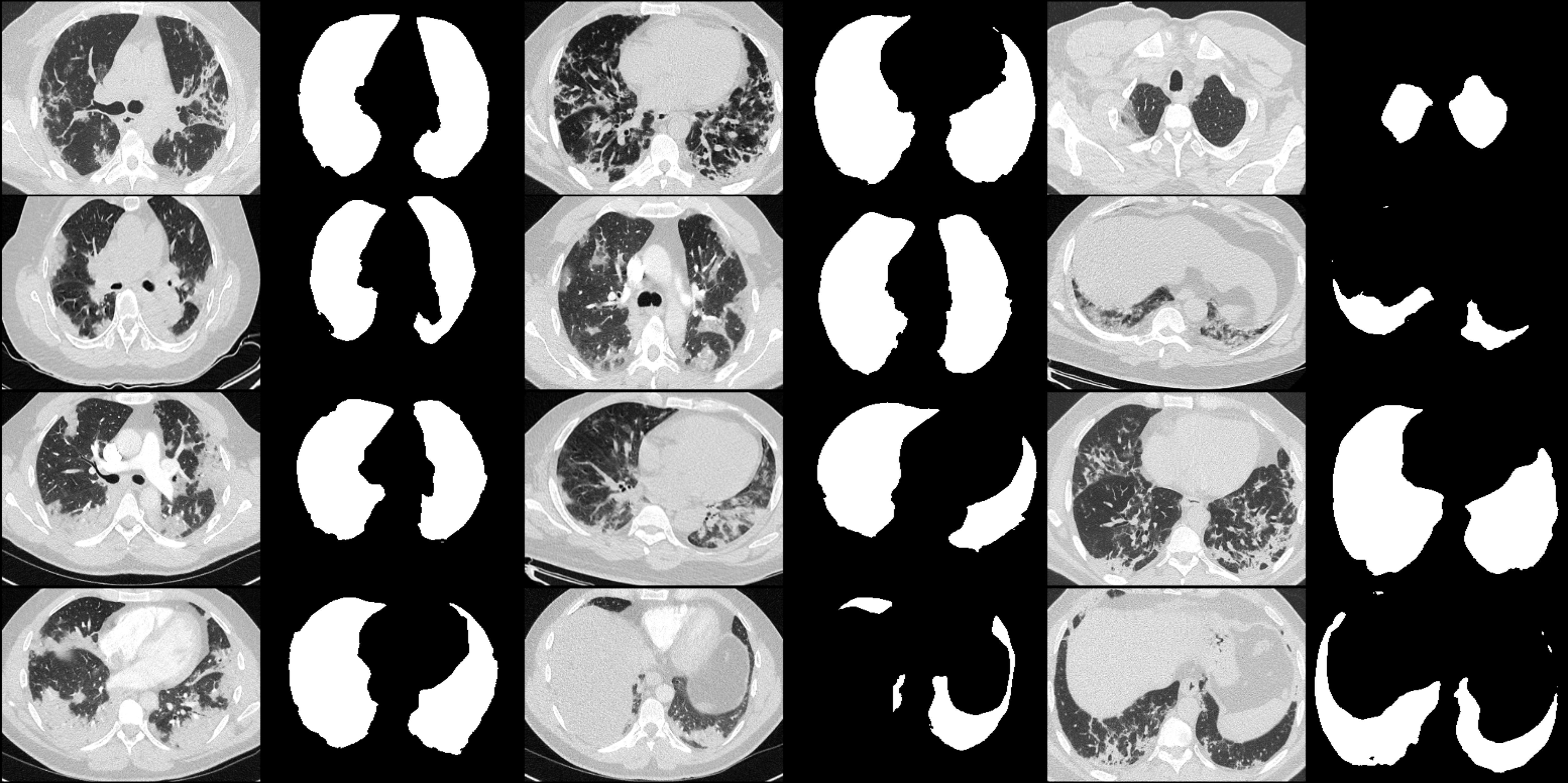}
\caption{Sample segmentation masks from the \textit{SILICOVILUNG} dataset obtained with the proposed CycleGAN+U-Net model.}
\label{fig:segmentation_silicovilung}

\end{figure}
 

\newpage
\subsection{Comparative qualitative analysis of segmentation models: our model vs. R231}

To benchmark our method against state-of-the-art approaches, we chose the widely recognized R231 model \cite{Hofmanninger_2020} from the Covid website, a prevalent tool in lung segmentation and a default extension in the 3D Slicer software, facilitating pathological lung segmentation. To ensure a fair comparison, we utilized the NIfTI DICOM files from the Covid challenge dataset for both the R231 model and our method. Given that our model is trained on grayscale images, we applied a windowing filter with a width of 1500 HU and a center of -500 HU. The performance metrics, in terms of Dice score, were computed for each method at patient level. 
\begin{figure}[htbp]

  \includegraphics[width=1\linewidth]{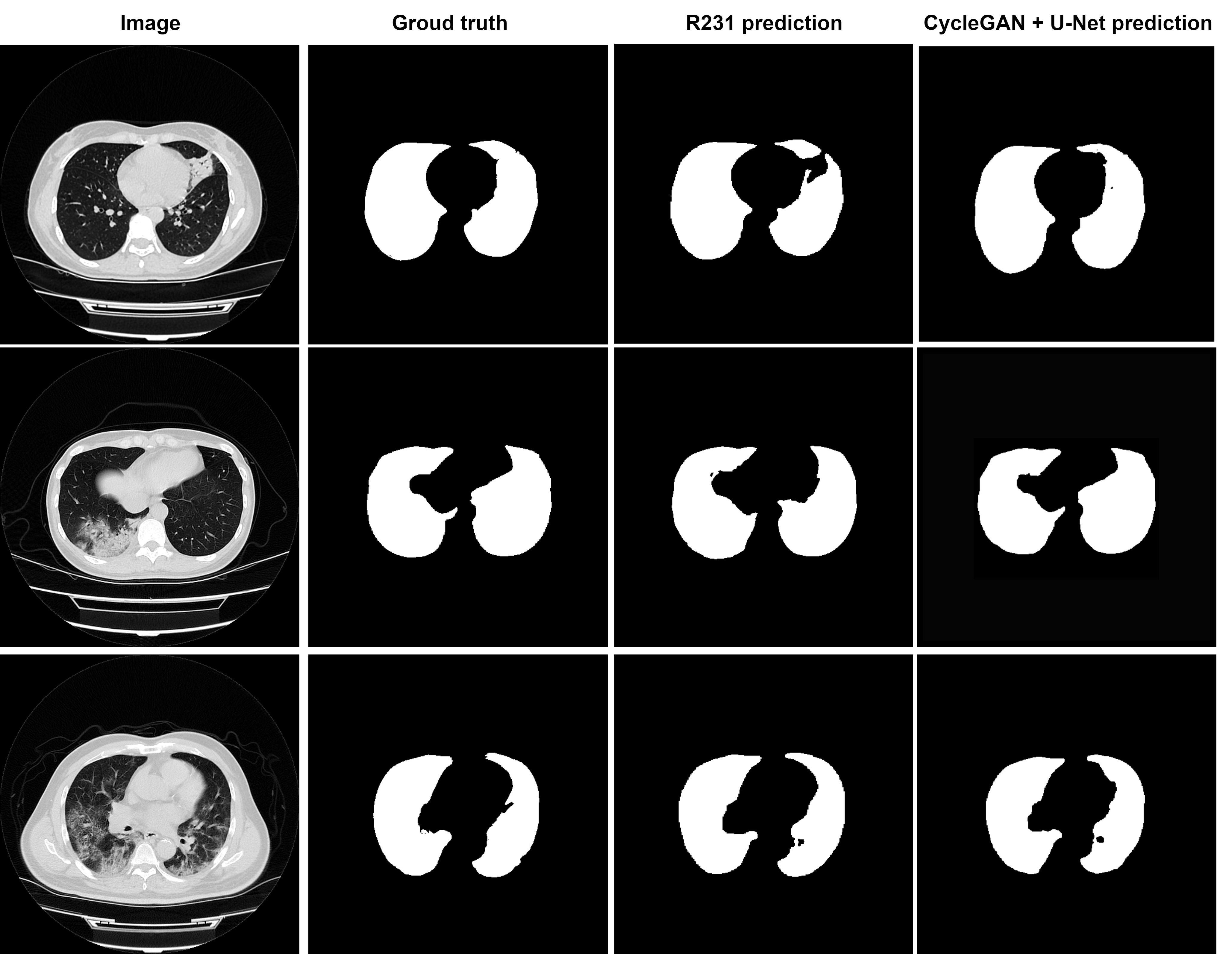}
  \caption{Qualitative comparison of segmentation outcomes for CycleGAN+U-Net and R231 models on the pathological subset-333 axial slices}
  \label{fig:results_segmentation_covid}
\end{figure}

\newpage
\section{Quantitative analysis}
\label{appendix:b}

\subsection{Comparative analysis of Dice scores for models on Covid-2019 dataset.}
\begin{table}[h!]
\centering
\caption{Comparative analysis of Dice scores for models tested on Covid-2019 dataset.}
\label{tab:comparison}
\resizebox{\textwidth}{!}{%
\begin{tabular}{lSccccc}
\toprule
\textbf{Patient ID} & \textbf{Pathology amount \%} & \multicolumn{5}{c}{\textbf{Dice Score}} \\
\cmidrule(lr){3-7}
& & \textbf{U-Net post-COVID} & \textbf{U-Net extended} & \textbf{V-Net} & \textbf{R231} & \textbf{CycleGAN+U-Net} \\
\midrule
A-0011 & 46.36 & 91.36 & 96.35 & 87.15 & 97.88 & 97.69 \\
A-0025 & 86.96 & 77.27 & 86.79 & 86.52 & 95.15 & 93.87 \\
A-0110 & 63.04 & 89.58 & 95.27 & 86.96 & 97.13 & 96.75 \\
A-0129 & 47.06 & 84.09 & 94.67 & 85.74 & 96.00 & 95.62 \\
A-0214 & 82.93 & 61.54 & 86.23 & 85.32 & 84.18 & 94.21 \\
A-0236 & 46.88 & 67.73 & 88.88 & 87.08 & 92.58 & 91.87 \\
A-0252 & 35.85 & 83.17 & 96.96 & 87.84 & 97.36 & 97.22 \\
A-0256 & 74.07 & 86.71 & 94.59 & 86.45 & 96.76 & 95.83 \\
A-0263 & 35.09 & 86.62 & 94.05 & 84.90 & 97.36 & 96.97 \\
A-0299 & 55.81 & 90.40 & 93.13 & 86.13 & 96.66 & 95.70 \\
A-0382 & 60.78 & 88.76 & 93.83 & 86.66 & 97.10 & 96.10 \\
A-0392 & 67.44 & 94.41 & 95.92 & 87.30 & 96.24 & 96.06 \\
A-0407 & 80.00 & 52.82 & 68.87 & 85.60 & 84.79 & 87.54 \\
A-0455 & 75.00 & 84.30 & 96.06 & 86.39 & 97.16 & 96.62 \\
A-0476 & 89.36 & 65.85 & 91.59 & 85.25 & 96.15 & 95.54 \\
A-0511 & 51.52 & 67.76 & 85.67 & 87.77 & 90.49 & 89.91 \\
A-0525 & 65.31 & 87.99 & 94.02 & 87.02 & 95.89 & 95.14 \\
A-0537 & 47.62 & 82.00 & 90.07 & 84.47 & 96.70 & 95.05 \\
A-0548 & 46.15 & 83.99 & 90.82 & 86.71 & 94.46 & 93.87 \\
A-0557 & 90.91 & 31.55 & 61.88 & 80.07 & 78.41 & 84.14 \\
A-0570 & 65.12 & 69.86 & 86.32 & 79.81 & 94.27 & 93.50 \\
A-0573 & 86.27 & 47.62 & 83.08 & 87.37 & 89.31 & 91.40 \\
A-0575 & 72.55 & 76.27 & 91.81 & 84.82 & 96.26 & 95.96 \\
A-0581 & 30.36 & 86.07 & 94.00 & 85.18 & 96.69 & 96.33 \\
A-0604 & 71.15 & 73.64 & 89.50 & 87.44 & 96.08 & 94.30 \\
A-0612 & 81.97 & 91.16 & 97.06 & 87.70 & 97.59 & 97.34 \\
A-0638 & 92.31 & 60.08 & 78.27 & 86.26 & 91.47 & 91.33 \\
A-0643 & 77.19 & 50.81 & 73.90 & 84.40 & 95.94 & 89.30 \\
A-0656 & 85.19 & 85.63 & 92.64 & 85.53 & 96.98 & 95.85 \\
A-0657 & 73.21 & 69.30 & 85.52 & 86.89 & 94.50 & 93.54 \\
A-0658 & 40.43 & 76.87 & 92.08 & 87.23 & 87.70 & 91.19 \\
A-0670 & 55.10 & 82.36 & 91.41 & 84.55 & 85.35 & 93.67 \\
A-0698 & 73.33 & 40.06 & 72.42 & 86.33 & 91.03 & 89.70 \\
\bottomrule
\end{tabular}}
\end{table}

\newpage
\subsection{Comparative analysis of Dice scores for models tested on ILD168 test set.}

\begin{table}[h!]
\centering
\caption{Comparative analysis of Dice scores for models on ILD168 test set.}
\label{tab:comparison_ild}
\resizebox{\textwidth}{!}{%
\begin{tabular}{
l 
S[table-format=2.1] 
S[table-format=1.4] 
S[table-format=1.4] 
S[table-format=1.4] 
S[table-format=1.4] 
}
\toprule
\textbf{Patient ID} & \textbf{Pathology amount \%} & \multicolumn{4}{c}{\textbf{Models}} \\
\cmidrule(lr){3-6}
& & \textbf{U-Net post-COVID} & \textbf{U-Net Extended} & \textbf{R231} & \textbf{CycleGAN+U-Net} \\
\midrule
PAT01 & 46.55 & 75.59 & 97.64 & 98.37 & 97.76 \\
PAT02 & 29.08 & 91.41 & 97.53 & 98.68 & 98.58 \\
PAT03 & 21.94 & 71.50 & 94.76 & 97.32 & 97.38 \\
PAT04 & 57.54 & 55.80 & 93.59 & 97.34 & 97.31 \\
PAT05 & 17.81 & 91.64 & 97.72 & 98.35 & 98.33 \\
PAT06 & 11.47 & 83.72 & 98.07 & 98.75 & 98.66 \\
PAT07 & 15.30 & 97.35 & 97.95 & 98.52 & 98.51 \\
PAT08 & 39.13 & 96.10 & 98.71 & 98.93 & 98.95 \\
PAT09 & 22.64 & 87.85 & 95.76 & 96.61 & 96.61 \\
PAT10 & 16.74 & 71.79 & 96.86 & 98.03 & 97.52 \\
PAT11 & 5.75 & 83.04 & 98.07 & 98.57 & 98.63 \\
PAT12 & 49.23 & 92.02 & 96.92 & 98.40 & 98.14 \\
PAT13 & 45.66 & 95.98 & 97.76 & 98.35 & 97.95 \\
PAT14 & 49.61 & 94.44 & 97.33 & 98.42 & 98.31 \\
PAT15 & 36.31 & 66.60 & 94.05 & 97.32 & 96.52 \\
PAT16 & 15.98 & 88.06 & 98.41 & 98.42 & 98.50 \\
PAT17 & 10.15 & 98.45 & 96.67 & 98.86 & 97.91 \\
\bottomrule
\end{tabular}}
\end{table}
\newpage
\subsection{Comparative analysis of Dice scores for models tested on Coronacases-Radiopedia dataset.}
The original dataset contains two subsets: coronacases and radiopedia cases. For the coronacases subset, the values are stored in 16-bit format, and we applied a windowing filter with a center of -500 HU and a width of 1500 HU to preprocess the data. However, in the case of the radiopedia subset, the values are already in 8-bit format. According to the official GitHub repository of the R231 model, the input must be in 16-bit HU values for accurate segmentation. Although the model provides an option `--noHU` for processing non-HU data, the authors caution that this option may not provide guaranteed results. Nevertheless, we tested the R231 model with the `--noHU` option on the radiopedia subset, but unfortunately, the predicted masks were all zeros. Consequently, we could not provide scores for the R231 model on the radiopedia cases due to this limitation.
\begin{table}[h!]
\centering
\caption{Comparative analysis of Dice scores for models tested on Coronacases-Radiopedia dataset. R231 could not be evaluated on the radiopedia dataset since it requires images in 16-bits Nifti format, whereas this dataset is provided as 8-bit grayscale images.}
\label{tab:comparaison_radiopedia}
\resizebox{\textwidth}{!}{%
\begin{tabular}{
    l 
    S[table-format=2.2] 
    S[table-format=1.4] 
    S[table-format=1.4] 
    S[table-format=1.4] 
    S[table-format=1.4] 
    S[table-format=1.4] 
}
\toprule
\textbf{Patient ID} & \textbf{Pathology amount \%} & \multicolumn{5}{c}{\textbf{Models}} \\
\cmidrule(lr){3-7}
 & & {\textbf{Unet post-COVID}} & {\textbf{Unet extended}} & {\textbf{V-Net}} & {\textbf{R231}} & {\textbf{CycleGAN+U-Net}} \\
\midrule
coronacases\_001 & 12.91 & 96.07 & 96.19 & 83.21 & 98.36 & 97.80 \\
coronacases\_002 & 4.15 & 96.45 & 97.37 & 86.04 & 98.24 & 97.92 \\
coronacases\_003 & 29.57 & 89.29 & 88.56 & 84.13 & 97.60 & 97.43 \\
coronacases\_004 & 1.40 & 98.05 & 98.01 & 83.15 & 98.65 & 98.26 \\
coronacases\_005 & 1.73 & 97.71 & 98.03 & 81.08 & 98.66 & 98.30 \\
coronacases\_006 & 2.92 & 96.43 & 97.30 & 84.86 & 98.65 & 98.29 \\
coronacases\_007 & 3.14 & 95.00 & 95.17 & 81.32 & 98.34 & 97.91 \\
coronacases\_008 & 7.58 & 95.42 & 96.25 & 89.04 & 98.59 & 98.15 \\
coronacases\_009 & 3.96 & 95.82 & 95.96 & 90.27 & 98.17 & 97.34 \\
coronacases\_010 & 17.71 & 90.98 & 92.59 & 80.40 & 97.68 & 95.59 \\
radiopaedia\_10\_85902\_1 & 3.47 & 86.26 & 94.93 & 78.04 & / & 97.45 \\
radiopaedia\_10\_85902\_3 & 3.24 & 82.70 & 96.46 & 80.48 & / & 98.14 \\
radiopaedia\_14\_85914\_0 & 25.46 & 94.82 & 95.43 & 82.93 & / & 97.57 \\
radiopaedia\_27\_86410\_0 & 10.35 & 72.14 & 94.87 & 85.37 & / & 96.71 \\
radiopaedia\_29\_86490\_1 & 0.01 & 85.41 & 94.12 & 87.82 & / & 96.81 \\
radiopaedia\_29\_86491\_1 & 0.88 & 95.00 & 97.66 & 90.26 & / & 97.82 \\
radiopaedia\_36\_86526\_0 & 2.79 & 98.24 & 98.24 & 91.99 & / & 98.64 \\
radiopaedia\_40\_86625\_0 & 59.37 & 86.97 & 93.49 & 88.97 & / & 96.04 \\
radiopaedia\_4\_85506\_1 & 21.01 & 90.26 & 94.41 & 92.00 & / & 97.20 \\
radiopaedia\_7\_85703\_0 & 18.65 & 89.84 & 93.01 & 92.05 & / & 98.00 \\
\bottomrule
\end{tabular}
}
\end{table}

\newpage 
\section{Detailed study on the impact of $\lambda$ values on model performance}
\label{appendix:c}
To optimize the $\lambda$ parameter associated with the L1 loss function in our model, we adjusted its coefficient from 1 to 10. This range corresponds to the maximum value of the cycle consistency loss defined in the CycleGAN official paper. However, we observed that when the $\lambda$ value was set to 2 or higher, the network less and less altered the lung content in terms of pathological texture generation. Conversely, a $\lambda$ value of 0, which indicates that the contribution of the L1 loss in the final loss calculation is removed, led the model to alter the lung shape, as illustrated in Figure \ref{fig:lambda_ablation}. Consequently, we determined the optimal $\lambda$ value to be 1, effectively balancing our model's ability to preserve the lung shape without compromising the generation of pathological textures.

As for the training specifics, we employed the Adam optimizer for both the generators and discriminators, with a fixed learning rate set to \(2 \times 10^{-4}\). In terms of data preprocessing, we applied z-score normalization to each dataset.

\begin{figure}[htbp]

  \includegraphics[width=\linewidth]{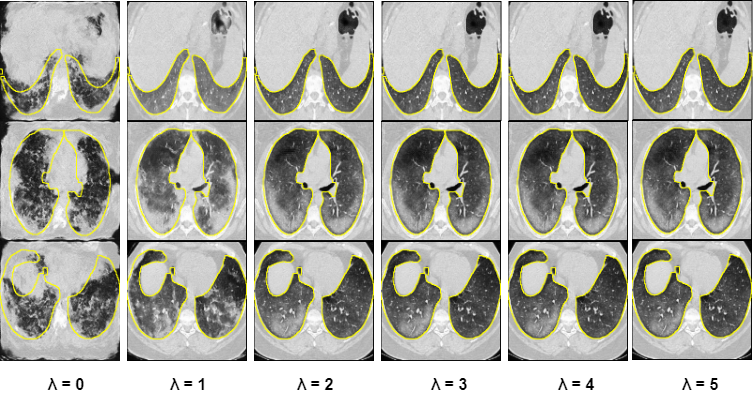}
 \caption{Qualitative comparison of model outputs across different $\lambda$ values weighting the $\mathcal{L}_{L1}(G, M)$ loss (Eq \ref{eq:Total_loss}) }
  \label{fig:ambda_ablation}
\end{figure}

\newpage
\section{Practical implementation }
\subsection{Practical implementation details of CycleGAN}
This appendix provides an overview of the practical aspects of implementing the CycleGAN model used in this study, inspired by the work of Aladdin Persson. Our implementation is based on PyTorch.

\subsection*{Codebase and modifications}

\begin{itemize}
    \item \textbf{Integration of testing phase:} We modified the code to include a testing phase at the end of each training epoch. This allows for continuous monitoring of the model's performance on unseen data throughout the training process.
    \item \textbf{Precision of computations:} Unlike some implementations that utilize mixed precision to accelerate training, our model strictly uses 32-bit floating-point.

\end{itemize}

\subsection*{Model optimization}
For optimizing our CycleGAN model, we adhered to the following settings, which align with those used in the original CycleGAN paper:

\begin{itemize}
    \item \textbf{Optimizer:} We employed the Adam optimizer, with a learning rate of $2 \times 10^{-4}$, consistent with the original implementation's recommendations.
    \item \textbf{Loss function parameters:} The lambda parameter for the L1 loss was set to 1, and for the cycle consistency loss, it was set to 10. These values are crucial for balancing the different components of the total loss function.
    \item \textbf{Batch size:} The model was trained with a batch size of 1. This setting was chosen to mirror the training conditions of the official CycleGAN, facilitating direct comparisons between our results and those reported in the literature.
\end{itemize}
\subsection{Practical implementation details of U-Net + CycleGAN model}

This appendix delves into the specifics of our approach in implementing the U-Net model augmented with CycleGAN for segmentation tasks. The implementation leverages PyTorch Lightning and MONAI, which facilitate efficient and clean code for medical image analysis.

\subsection*{Model architecture and enhancements}
Our model builds on the foundational architecture of the original U-Net, incorporating key enhancements to improve performance and mitigate overfitting:

\begin{itemize}
    \item \textbf{Batch normalization:} Batch normalization layers were added to the  the U-Net architecture.
    \item \textbf{Dropout:} To prevent overfitting, dropout layers were integrated into the decoder with a probability of 0.2. 
    \item \textbf{Gradient clipping:} The training process also utilized gradient clipping with a norm value set to 1. This technique prevents the exploding gradients problem by limiting the value of gradients to a defined range.
\end{itemize}

\subsection*{Optimization strategy}
The optimization of the model was carefully designed to ensure robust learning dynamics:

\begin{itemize}
    \item \textbf{Optimizer:} We employed the SGD optimizer, with an initial learning rate of 0.01, momentum of 0.9, and a weight decay of 0.01. These parameters were chosen to promote a steady and controlled update of model weights.
    \item \textbf{Cyclical learning rate:} A cyclical learning rate strategy was adopted, with the base and maximum learning rates set between 0.01 and 0.1, respectively. The learning rate varies in an exponential range mode, allowing the model to explore the parameter space more effectively and avoid local minima.
\end{itemize}

\subsection{Experimental setup}
All experiments were conducted on an NVIDIA RTX-A6000 device with cuda version 11.7

\end{document}